# Spectral beam combining of Yb-doped fiber lasers using wavelength dependent polarization rotators and polarization beam combiners

B. S. Tan[1*], P. B. Phua[1,2] and R. F. Wu[1]

[1]*DSO National Laboratories, 20, Science Park Drive, S118230, Republic of Singapore*
[2]*Nanyang Technological University, 50, Nanyang Avenue, S639798, Republic of Singapore*
*\* Corresponding Author: tbengsin@dso.org.sg*

**Abstract**

We propose, for the first time, a robust spectral beam combining scheme using wavelength dependent polarisation rotators and polarization beam combiners. We successfully demonstrated the concept for two Yb-doped fiber lasers at 1064nm and 1092nm up to a total input power of 90W. The results reveal a very good combining efficiency and the potential for scaling to high power operations in this method of beam combining.

*OCIS* codes: 140.3510, 140.3300, 140.3290. 230.5440

A key challenge of building high brightness and high power lasers centers on the effective dissipation of heat generated in the laser medium. The unwanted heat often led to thermo-optical distortions which resulted in laser output of poor beam quality. To increase the laser power whilst maintaining the beam quality i.e. brightness scaling, both coherent and wavelength beam combining have been actively explored by many groups [1~5] in recent years. In coherent beam combining, several diffraction limited beams of the same wavelength are phase-locked so that their fields sum up coherently in the near or far field. The phase locking can be achieved either passively through an intra-cavity arrangement [1, 2] or actively via an external phase control system [3]. The former has the advantage of a relatively simple architecture though its scaling potential remains doubtful [6]; with the latter, 19kW of combined power with near diffraction limited quality has been demonstrated [3], the brightest cw solid-state laser ever reported, but stringent phase control requirement for the combining arms results in a more complex setup. Wavelength beam combining, on the other hand, involves laser arms of slightly different wavelengths which are combined through one or more dispersive elements. Due to the less stringent requirements for the laser properties [6], the schemes for wavelength combining are generally simpler than those of coherent combining. With output of Yb-doped fiber lasers reaching kW level [7~9] and offering broad emission bandwidths ranging from 1030nm to 1100nm, wavelength combining schemes employing Yb-doped fiber lasers are therefore particularly attractive for high power lasers applications which are not wavelength specific. Dispersive elements such as gratings [4, 5] and dichroic mirrors [10] have been employed with some success but the combining efficiencies, limited by the spectral selectivity and diffraction losses for the dichroic mirrors and grating respectively, and the power handling of these elements seem to prevent further power scaling.

In this Letter, we propose a highly efficient method of wavelength beam combining, using a wavelength dependent polarization (WDP) rotators and polarization beam combiners. The setup is simple and employs standard optical elements which allow very robust operation under high power operations. The basic idea of our beam combining scheme is explained with two beams as illustrated in Figure 1. Two linearly and orthogonally polarized beams of different wavelengths are combined spatially through a polarization beam combiner. The combined beam is sent through a WDP rotator which transforms the orthogonally polarized spectral lines into the same linearly polarized state. The use of a polarization beam combiner notwithstanding, the method is spectral and not polarization beam combining. This is clear when one considers that the beam above can be further combined with a spectrally identical but orthogonally polarized beam through a polarization beam combiner.

The key element in this scheme is the WDP rotator. It was first introduced in reference [11] where it was used to transform the orthogonal polarizations of the signal and idler beams of a near degenerate type II KTP optical parametric oscillator into a common linearly polarized state. It comprises of three optical elements: (i) a birefringent plate, (ii) a broadband quarter-wave plate and (iii) a broadband half-wave plate. With the principal axes of the birefringent plate aligned at 45º to the horizontal direction, a beam whose two spectral lines are orthogonally polarized in the horizontal and vertical directions attains a common state of polarization (SOP) for its two spectral components when the thickness of the birefringent plate satisfies $l = (2m_{ij} + 1)l_{ij}$ where $m_{ij} = 0, 1, 2, 3, \ldots$. The zero-order (when $m_{ij} = 0$) thickness $l_{ij}$ is given by [11]

$$l_{ij} = \frac{1}{2}\left|\Delta n_i / \lambda_i - \Delta n_j / \lambda_j\right|^{-1} \qquad (2)$$

Here, $\Delta n_i$ and $\Delta n_j$ refer to the birefringence at wavelengths $\lambda_i$ and $\lambda_j$ respectively. The common SOP is at this stage, generally, elliptical and the appropriately aligned quarter-wave and half-wave plates further transform the SOP to one which is linearly polarized in the desired orientation.

A straightforward way of extending this method of beam combining using WDP rotators and polarization beam combiners beyond two beams is to sequentially add beams of different wavelengths. Consider the addition of a linearly polarized beam of wavelength $\lambda_3$ to a linearly polarized beam of spectral lines $\lambda_1$ and $\lambda_2$. The beams are spatially combined using a polarization beam combiner. Further additions of beams of other wavelengths require a prior transformation of the polarizations of the three spectral lines into a common linearly polarized state using a WDP rotator as in the case of two beams. Specifically the thickness of the birefringent plate in this WDP rotator must satisfy $l = (2m_{13} + 1)l_{13} = (2m_{23} + 1)l_{23}$. To meet this criteria is possible but $m_{13}$ and $m_{23}$ may be too large (depending on $l_{13}$ and $l_{23}$), possibly resulting in an impractical size for the birefringent plate. This problem worsens with the addition of further beams. The difficulty can be overcome, however, by imposing a periodicity on the wavelengths of the beams. With wavelengths $\lambda_i \sim \lambda_j \sim \lambda$ such that $\Delta n_i \sim \Delta n_j \sim \Delta n$, equation (2) can be simplified to

$$l_{ij} = \lambda^2 /(2\Delta n \Delta\lambda_{ij}) \quad (3)$$

In this approximation, $l_{ij}$ is dependent on $\Delta\lambda_{ij} = |\lambda_i - \lambda_j|$, the wavelength difference between the two spectral lines. Now consider the problem of splitting a linearly polarized multi-wavelength beam, whose $2^n$ spectral lines, are separated by $\Delta\lambda$, as illustrated in Figure 2. When the beam passes through a WDP rotator with its birefringent plate thickness $l_{ij}$ satisfying equation (3) with $\Delta\lambda_{ij} = \Delta\lambda$, the spectral lines of consecutive wavelengths become orthogonally polarized and the beam is separated by the polarization beam combiner into two orthogonally polarized beams. Each beam has $2^{n-1}$ spectral lines separated apart by $2\Delta\lambda$. With a second pair of WDP rotator (with $l_{ij}$ corresponding to $\Delta\lambda_{ij} = 2\Delta\lambda$) and polarization beam combiner, the two beams can be further separated into four beams, each with $2^{n-2}$ spectral lines separated apart by $4\Delta\lambda$. Thus, by passing the beam through several stages each consisting of a WDP rotator (with an appropriate $l_{ij}$ corresponding to the wavelength separation $\Delta\lambda_{ij}$ at each stage) and a polarization beam combiner, the beam can be split into its individual spectral components. The reversibility of the propagation means that the same architecture can be employed to combine multiple monochromatic beams whose wavelengths obey this periodic structure. This periodicity requirement is easily met with Yb-doped fibers given their broad and continuous spectrum.

    With the scheme just described, it is clear that the efficiency of our scheme depends on how well the beams can be spatially combined at the polarization beam combiners which, in turns, depends on how closely matched the polarizations of the

spectral lines are after passing through WDP rotators. A perfect match is obtained when the birefringent plate in the WDP rotator has thickness $l = (2m_{ij}+1)l_{ij}$ but this may not be the case for two reasons. Firstly, the precision required for $l_{ij}$ may be far more stringent that what the fabrication of the birefringent plates can achieve. Secondly, all laser beams have finite linewidths whereas the thickness $l_{ij}$ is optimized for a specific wavelength. For errors $\delta l$ in the design thickness $l = (2m_{ij}+1)l_{ij}$ of the birefringent plate and $\delta\lambda$ in the wavelength difference $\Delta\lambda_{ij}$ of the two beams, the angle separation $\delta\theta$ of the linear polarizations of the combined beam after passing through the WDP rotator, from theoretical considerations, is $(\delta l/l_{ij})\pi/2$ and $(2m_{ij}+1)(\delta\lambda/\Delta\lambda_{ij})\pi/2$ respectively. The optimized combining efficiency at a polarization beam combiner for two beams whose spectral lines' polarizations are separated by $\delta\theta$ is given by $(1+\cos\delta\theta)/2$. Efficiency higher than 95% can be obtained provided $\delta\theta$ is kept under 20º. This serves as a guideline on the upper limits of both $\delta l$ and $\delta\lambda$ for a good combining efficiency.

Our experimental setup is illustrated in Figure 3. We used Yb-doped fiber lasers (from IPG Photonics, Germany) as our two laser sources; one is a 100W rated oscillator (PYL-100) which emits a non-polarized beam of wavelength near 1092nm while the other is a master oscillator (YLD-0.01) power amplifier (YAR-47) which emits a non-polarized beam of output power up to 50W at 1064nm. Two thin film plate polarizers (TFP A and TFP B) were used to obtain the vertically and horizontally polarized beams. In addition, TFP A also acted as a polarization beam combiner to spatially combine the orthogonally polarized 1064nm beam and 1092nm beam. For our WDP rotator, we employed a y-cut quartz plate with a thickness of 6.28μm as the birefringent plate followed by a zero-order quarter-wave quartz plate and a zero-order half-wave quartz plate. All three optical elements are AR coated at 1064nm and the measured transmission loss for the combined beam through the WDP rotator was negligible as expected. Crystal quartz was chosen as the substrate for the wave plates primarily due to its known capacity to handle high power. Commercially available quartz wave plates with AR coatings at 1μm have damage threshold rated up to 1MWcm$^{-2}$ of cw power. The other advantage of using crystal quartz is that the birefringence of 0.087 at 1μm is able to meet the broad emission bandwidth of Yb-doped fibers (approximately 60nm) and provide a good spectral resolution of 1nm for wavelength selection with reasonable thicknesses. Using equation (2), the thickness $l_{ij}$ for a wavelength difference $\Delta\lambda_{ij}$ of 1nm and 60nm are 60mm and 1mm respectively. Quartz plates with a thickness within this 1mm to 60mm range are easy to fabricate and to work with experimentally. For our experiment, the zero-order thickness $l_{ij}$ corresponding to the two wavelengths is 2.17mm so the thickness of our birefringent quartz plate is 230μm less than the first order ($m_{ij} = 1$) thickness of 6.51mm giving a $\delta l$ less than 11% of $l_{ij}$. The FWHM linewidth of the 1092nm laser source when operated near its maximum rated power of 100W is approximately 2nm whereas the linewidth of the 1064nm laser source stays constant at 0.1nm independent of the operating power. The effect of the linewidth on the beam combining efficiency when the two lasers operate at full power is thus dominated by the 1092nm source and one can approximate this effect of the linewidth on the combining efficiency by taking $\delta\lambda$ to be 1nm which is less than 4% of $\Delta\lambda_{ij}$. From our earlier analysis, the angle separation $\delta\theta$ between the polarizations of the two spectral components of the combined beam is approximately 10º for both $\delta l$ and $\delta\lambda$ so a very high combining efficiency is expected. The combined beam was sent through a diagnostic thin film plate polarizer (TFP C) to verify this. The spectrum of the combined beam, measured using an optical spectrum analyzer, with both lasers running near their maximum capacities is shown in Figure 4.

With up to 90W (44W and 46W from the 1064nm and 1092nm linearly polarized beams respectively) of spatially combined power, the transmissions of the combined beam was 83% when horizontally polarized and 2% when vertically polarized. The direction of the polarization of the combined beam is fixed by the orientation of the half-wave plate in the WDP rotator. These transmission figures were in fact fairly independent of the input power, in agreement with our theoretical analysis which suggests that the linewidth of the 1092nm source should have little effect on the performance of the WDP rotator. The low transmission of the combined beam when horizontally polarized was primarily due to the poor performance of thin film plate polarizers for p-polarized beams near 1092nm. We measured the transmissions of p-polarized beams at 1092nm to be 74% in contrast to the 95% for p-polarized beams at 1064nm. Taking these transmission figures into account, the polarizations of the two spectral lines in the combined beam were therefore very close as predicted. The combining efficiency for our setup was thus limited by the thin film plate polarizers which have polarizing bandwidths of only 5nm. With the use of high energy broadband polarizers with polarizing bandwidths matching the emission bandwidth of Yb-doped fiber in place of the thin film plate polarizers, the true potential of the scheme can be realized and high combining efficiencies of more than 95% is achievable.

In summary, we have proposed, for the first time, a method of beam combining using WDP rotators and polarization beam combiners. The method employs standard optical elements which enable very robust operation at high power with the imposition of a periodicity on the wavelengths being the only restriction. We combined two beams of wavelengths 1064nm and 1092nm up to 90W of input power to analyze the scheme's potential. Taking into account the performance of the thin film plate polarizer at 1092nm, the results agree well with our theoretical analysis which has revealed this method of beam combining to be highly efficient, and insensitive to both fabrication (for the WDP rotator) and spectral selectivity (for the laser sources) issues. With the use of high energy broadband polarizers as polarization beam combiners and highly efficient Yb-doped fibers, capable of providing both the spectral bandwidth and periodicity required for this method, as laser sources, we believe this spectral beam combining scheme has tremendous potential for scaling to high power.

The authors would like to acknowledge Dr Teo Kien Boon and Dr. Lai Kin Seng for their support and encouragement in this work.

**References**

[1] V. Eckhouse, A. A. Ishaaya, L. Shimshi, N. Davidson, and A. A. Friesem, Opt. Lett., **31**, 350 (2006)

[2] M. Minden, H. Bruesselbach, J. Rogers, D. C. Jones and M. Mangir, CLEO Europe, **TFII2-2-WED** (2005)

[3] G. D. Goodno, H. komine, S. J. McNaught, S. N. Weiss, S. Redmond, W. Long, R. Simpson, E. C. Cheung, D. Howland, P. Epp, M. Weber, M. Mcclellan, J. Sollee, and H. Injeyan, Opt. Lett., **31**, 1247 (2006).

[4] S. J. Augst, A. K. Goyal, R. L. Aggarwal, T. Y. Fan and A. Sanchez, Opt. Lett., **28,** 331 (2003)

[5] Igor V. Ciapurin L. B. Glebov, L. Glebova, V. I. Smirnov, E. V. Rotari, Proc. SPIE, **4974**, 210 (2003)

[6] T. Y. Fan, IEEE J. Sel. Top. Quant. Elect., **11**, 567 (2005)

[7] A. Liem, J. Limpert, H. Zellmer, A. Tünnermann, V. Reichel, K. Mörl, S. Jetschke, S. Unger, H. Müller, J. Kirchof, T. Sandrock, A. Harschak, CLEO USA, **CPDD2** (2004)

[8] Y. Jeong, J. Sahu, D. Payne, J. Nilsson, Opt. Exp., **12**, 6088 (2004)

[9] V. Gatponsev, D. Gapontsev, N. platonov, O. Shkurikhin, V. Fomin, A. Mashkin, M. Abramov, S. Ferin, CLEO Europe, **CJ1-1-THU** (2005)

[10] K. Nosu, H. Ishio, and K. Hashimoto, Electron. Lett., **15**, 414 (1979)

[11] P.B. Phua, B.S.Tan, R. F. Wu, K. S. Lai, Lindy Chia, and Ernest Lau, Opt. Lett., **31**, 489 (2006)

**Figure Captions**

Figure 1: Schematic diagram of a spectral beam combining scheme for two linearly polarized beams of wavelengths $\lambda_1$ and $\lambda_2$ using a polarization beam combiner (PBC) and a WDP rotator which comprises of a birefringent plate, a quarter-wave plate and a half-wave plate.

Figure 2: Schematic diagram of architecture, consisting of pairs of a WDP rotator and a polarization beam combiner (PBC), for splitting a multiple-wavelength, linearly polarized beam into its individual spectral lines. Bold arrows represent horizontally polarized spectral lines; dashed arrows represent vertically polarized spectral lines; thickness of each birefringent element in WDP rotator depends on wavelength separation $\Delta\lambda_{ij}$ (indicated in parentheses).

Figure 3: Schematic diagram of the experimental setup for combining two fiber laser sources at 1064nm and 1092nm. TFP: thin film plate polarizer; OSA: optical spectrum analyzer.

Figure 4: Measured spectrum for combined beam using an optical spectrum analyzer.

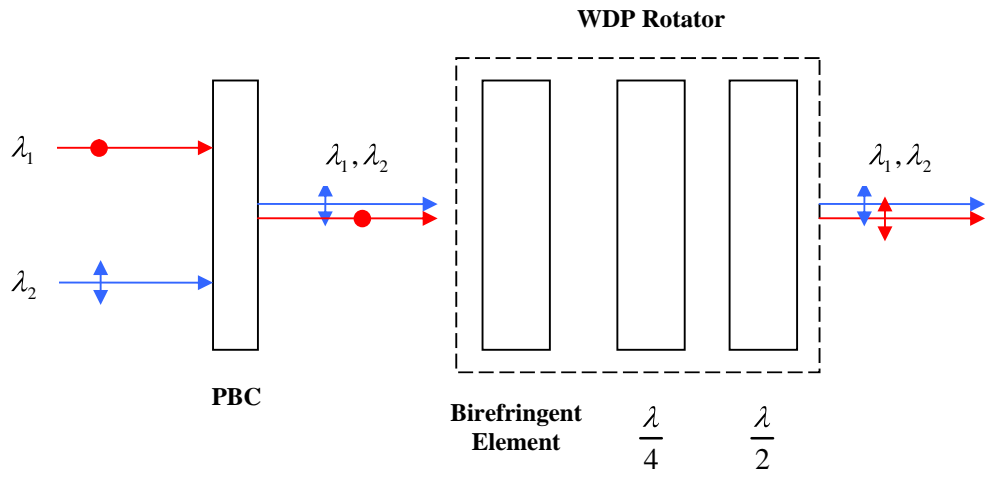

**Figure 1**

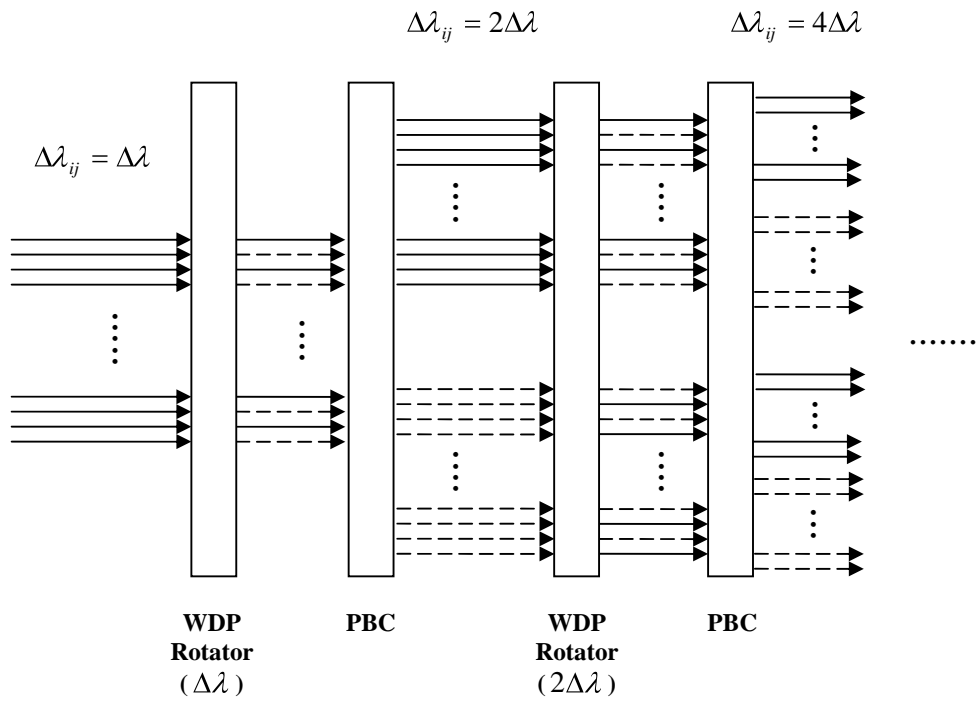

**Figure 2**

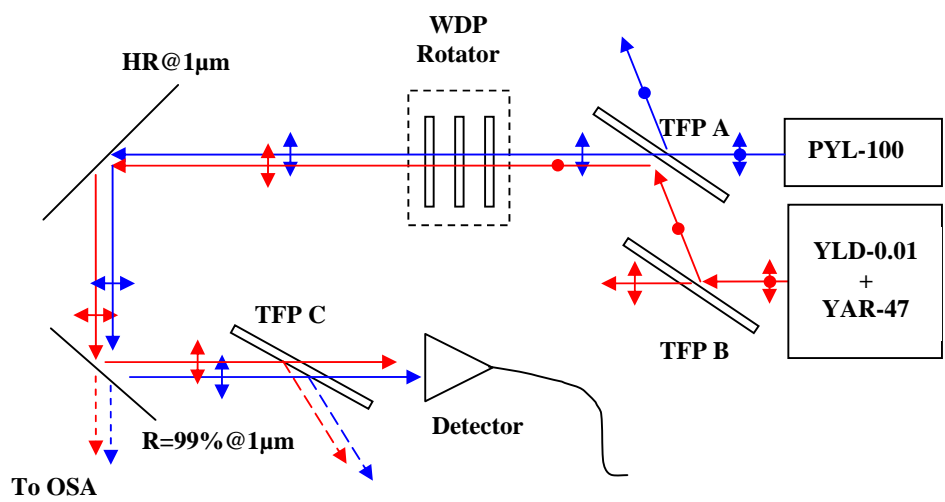

**Figure 3**

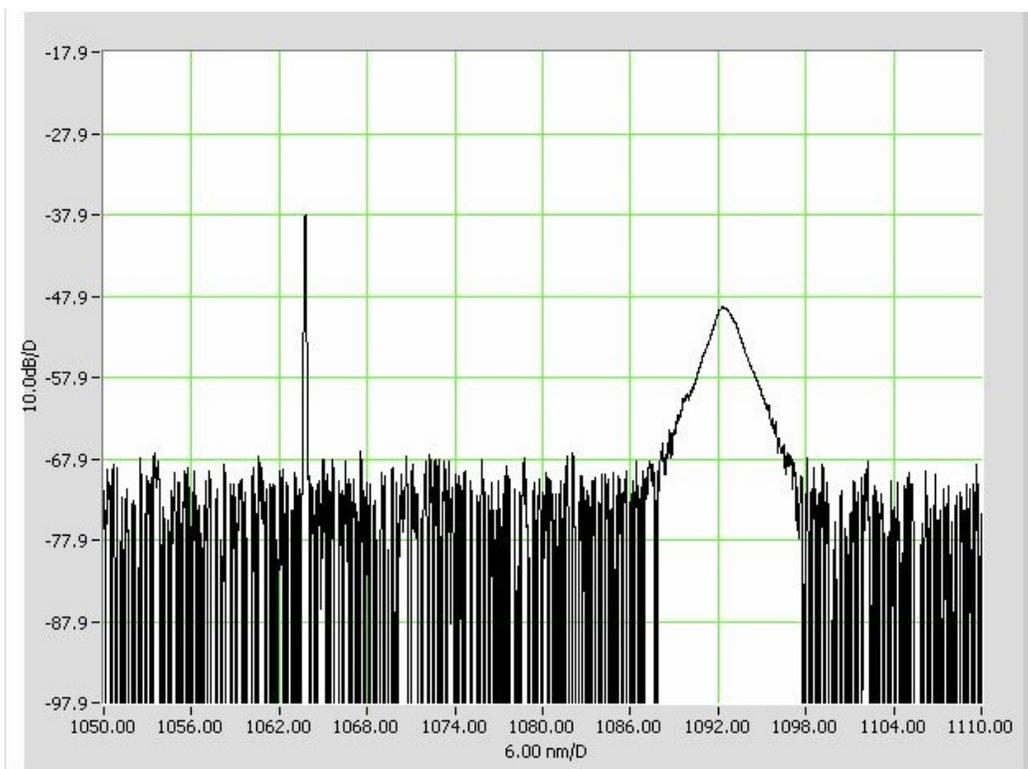

**Figure 4**